\documentclass[manuscript]{aastex}

\usepackage{graphicx}
\usepackage{dcolumn}
\usepackage{paralist}
\usepackage{comment}
\usepackage{graphicx,epsfig}
\usepackage{multirow}
\usepackage{wrapfig}
\usepackage{float}

\slugcomment{To be sumitted to AJ.}

\shorttitle{Ground-based NEA Astrometry}
\shortauthors{C.Zhai et al.}

\def\beq{\begin{equation}}
\def\eeq{\end{equation}}
\def\beqa{\begin{eqnarray}}
\def\eeqa{\end{eqnarray}}
\def\as{^{\prime\prime}}
\def\am{^\prime}
\begin{document}
\title{Accurate Ground-based Near-Earth-Asteroid Astrometry using Synthetic Tracking}

\author{Chengxing Zhai,
Mike Shao, Navtej Saini, Jagmit Sandhu, William M. Owen, Thomas A. Werne, Todd A. Ely,  Joseph Lazio, Tomas J. Martin-Mur, Robert A. Preston, Slava G. Turyshev,
}
\affil{Jet Propulsion Laboratory, California Institute of Technology, \\
4800 Oak Grove Dr, Pasadena, CA 91109}
\email{chengxing.zhai@jpl.nasa.gov}
\and
\author{Philip Choi, Adam W. Mitchell, Kutay Nazli, Isaac Cui}
\affil{Department of Physics, Pomona College, Claremont, CA, 91711}
\and
\author{Rachel M. Mochama}
\affil{Department of Physics, Scripps College, Claremont, CA, 91711}

\begin{abstract}

Accurate astrometry is crucial for determining orbits of near-Earth-asteroids (NEAs) and therefore better tracking them.
This paper reports on a demonstration of 10 milliarcsecond-level astrometric precision on a dozen NEAs using the Pomona College 40 inch telescope, at the JPL's Table Mountain Facility. 
We used the technique of synthetic tracking, in which many short exposure (1 second) images are acquired and then combined in post-processing to track both target asteroid and  reference stars across the field of view. 
This technique avoids the trailing loss and keeps the jitter effects from atmosphere and telescope pointing common between the asteroid and reference stars,  resulting in higher astrometric precision than the 100 mas level astrometry from traditional approach of using long exposure images.
Treating our synthetic tracking of near-Earth asteroids as a proxy for observations of future spacecraft while they are downlinking data via their high rate optical communication laser beams,
our approach shows precision plane-of-sky measurements can be obtained by the optical ground terminals for navigation. 
We also discuss how future data releases from the Gaia mission can improve our results.

\end{abstract}

\section{Introduction}
\begin{figure}[ht]
\epsscale{1}
\plotone{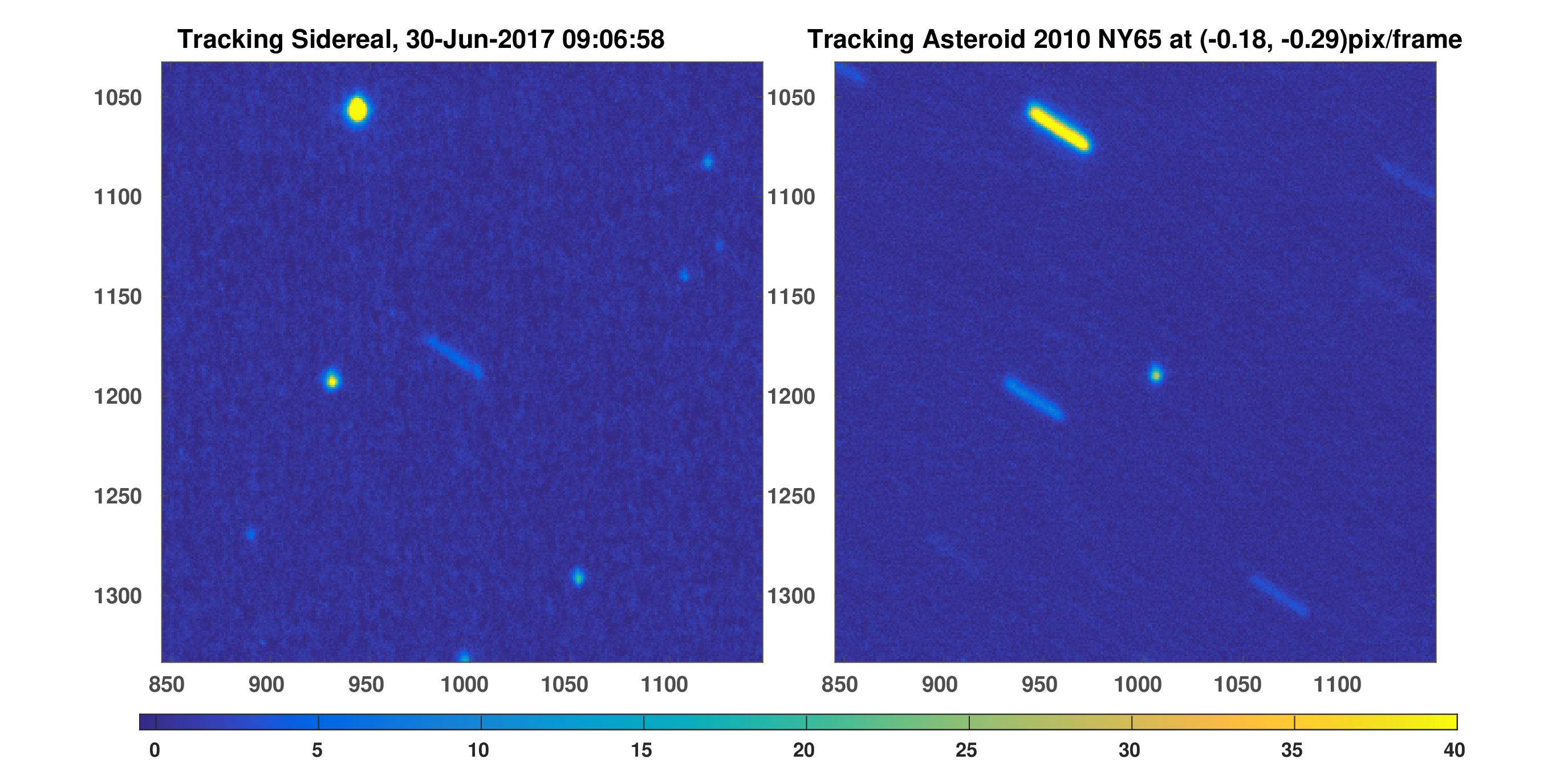}
\caption{Synthetic tracking allows integrating short-exposure frames in the post-processing to tracking the sidereal (left) and asteroid (right) by shifting the frames with the corresponding amount of motion.\label{SynTrack}}
\end{figure}

Near-Earth-Asteroid (NEA) observations and characterization is crucial for protecting our planet.
Following congressional directions, NASA has been actively detecting, tracking and characterizing potentially hazardous asteroids and comets that could approach the Earth. 
While the impact of a big asteroid could have catastrophic consequences,
in view of the damage caused by the incident of the Chelyabinsk meteor \citep {Brumfiel2013} and the fact that the frequency for smaller asteroids to impact earth is much higher
than that of larger asteroids \citep{NRC2010}, it is important to watch 
for any potential threats from asteroids larger than 10 meter. 
Larger asteroids are brighter and thus easier to detect. To efficiently detect small asteroids, we have developed the synthetic tracking technique which replaces
long exposure images with multiple short exposure images and integrates the images in post-processing to detect faint asteroids \citep{Shao2014, Zhai2014}.
Synthetic tracking is enabled by the modern sCMOS cameras that can take megapixel frames at rate faster than 10Hz yet only introduce low read noise at 1-2e$^-$ per read.%
\footnote{See \url{http://www.andor.com/scientific-cameras/neo-and-zyla-scmos-cameras}, and \url{https://www.photometrics.com/products/scmos/} for more information.}%
Synthetic tracking avoids streaked images, thus improves signal-to-noise ratio (SNR) for detection as shown in Fig.~\ref{SynTrack} where we display
the images taken by our instrument for tracking both the sky (left) and the asteroid 2010 NY65. The degraded SNR due to trailing loss is obvious because the asteroid appears much 
brighter in the right image compared with the left one and the faint stars in the left image can barely be seen in the right image.
Going to space, a constellation of SmallSats carrying telescope with synthetic tracking capability can be a very economic way to speed up the survey of NEAs \citep{Shao2017}.

The synthetic tracking technique also yields more accurate astrometry than the detection
with streaked images of asteroids \citep{Zhai2014} in addition to higher detection sensitivity.
Synthetic tracking successfully avoids three disadvantages of doing astrometry with streaked images: 1)
degraded SNR from trailing loss; 2)  poor sensitivity in centroiding along the streak due to non-compact PSF;
3) the errors due to atmospheric and telescope pointing jitter are no longer common between the moving target and reference stars.
Using synthetic tracking,  we can achieve astrometry accuracy for NEAs comparable to stellar astrometry
because we can track both the NEAs and reference stars in post-processing; the target and reference objects can be treated the same way.
This is significant in view of that ground-based stellar astrometry was able to reach 1 mas accuracy more than a decade ago\citep{Pravdo2004,Henry2009}.


The potential of highly accurate astrometry from using synthetic tracking is extremely valuable for cataloging the
discovered NEAs because currently a large number of observed NEAs are lost subsequently
due to inaccurate orbit determination \citep {Blair2002}.
Accurate astrometry also produces more accurate future orbital paths for close Earth approaches and thus more reliable probabilities of impacting earth.
Another application of accurate ground-based astrometry is for optical navigation.
The future of deep space high data rate communications is likely to be optical communications,
such as the Deep Space Optical Communications package that is part of the baseline payload for the planned Psyche Discovery mission to the Psyche asteroid%
\footnote{See \url{https://www.nasa.gov/directorates/spacetech/tdm/feature/Deep\_Space\_Communications}.}.%
Viewing asteroids as proxies for the future spacecraft that carry optical communication devices for higher data rate,
our accuracy can also serve as a metric of performance in measuring the spacecraft position in the plane of sky for optical navigation.

With the Gaia's Data Release 1 (DR1) catalog \citep{Gaia2016}, we are able to achieve better than 50 milli-arcsecond (mas) accuracy for most of the NEAs we observed 
during summer 2017 using an Andor's Zyla 5.5 sCMOS camera %
\footnote{See a description of technical
capabilities of the Andor's Neo and Zyla sCMOS cameras at: {\url
http://www.andor.com/pdfs/literature/Andor\_sCMOS\_Brochure.pdf}}
on the Pomona College's 40 inch telescope at the Table Mountain Facility.
For brighter asteroids, our best accuracy is about 10 mas with integration time of 100 seconds,
limited mainly by the chromatic distortion effect due to refractive optical elements in our system.
This paper reports our method and results from our instrument using synthetic tracking and Gaia DR1 catalog to demonstrate the potential of achieving accuracy for NEA astrometry
much higher than the current state-of-the-art of 120 mas from Pan-STARRS survey telescope \citep{Veres2017} .

\section{Instrument}
Pomona College's 40 inch telescope at the Table Mountain Facility (TMF) is a Cassegrain telescope with about 1 m size primary mirror at f/2.
With a 30 cm secondary mirror, we get an imaging system of focal length 9.6 meter.
An Andor Zyla 5.5 sCMOS detector is put behind the relay optics to have an effective system of f/2.8.
The 6.5 $\mu$m pixel at the effective focal length of 2.8 m gives us a plate scale of 0.45$\as$ per pixel, enabling a critical sampling of point-spread-function (PSF)
for typical seeing of $2\as$ at the TMF.
The array size is 2560$\times$2160, giving a field of view (FOV) of $19\am{\times}16\am$.
The sCMOS camera can run up to 100 frames per second with about 1.4e$^-$ read noise, suitable for observing very fast moving objects.
Using synthetic tracking, we would like to take frames at a rate so that the moving object does not streak compared with the size of the PSF. 
Our default frame rate is 1Hz, which is sufficient for most of the NEAs at seeing of 2$\as$. Note that even for the darkest night at TMF with sky
flux $\sim$ 20.5 mag per square arcsecond, we are still limited by sky background noise at 1Hz.
We took dark frames at 1 Hz and used twilight flat field measurements to estimate flat field response for calibration.

\section{Data Reduction}
A single data set of synthetic tracking contains multiple short exposure frames, which we call a data cube in the sense
that there is an extra time dimension in addition to the camera frame's row and column dimension.
The main task of data reduction is to estimate the centroid of  both the asteroid and reference stars in the field using the data cube and
solve for the sky position of the asteroid based on identified reference stars.

 \subsection{Overview of data processing}
\begin{figure}[ht]
\epsscale{0.8}
\plotone{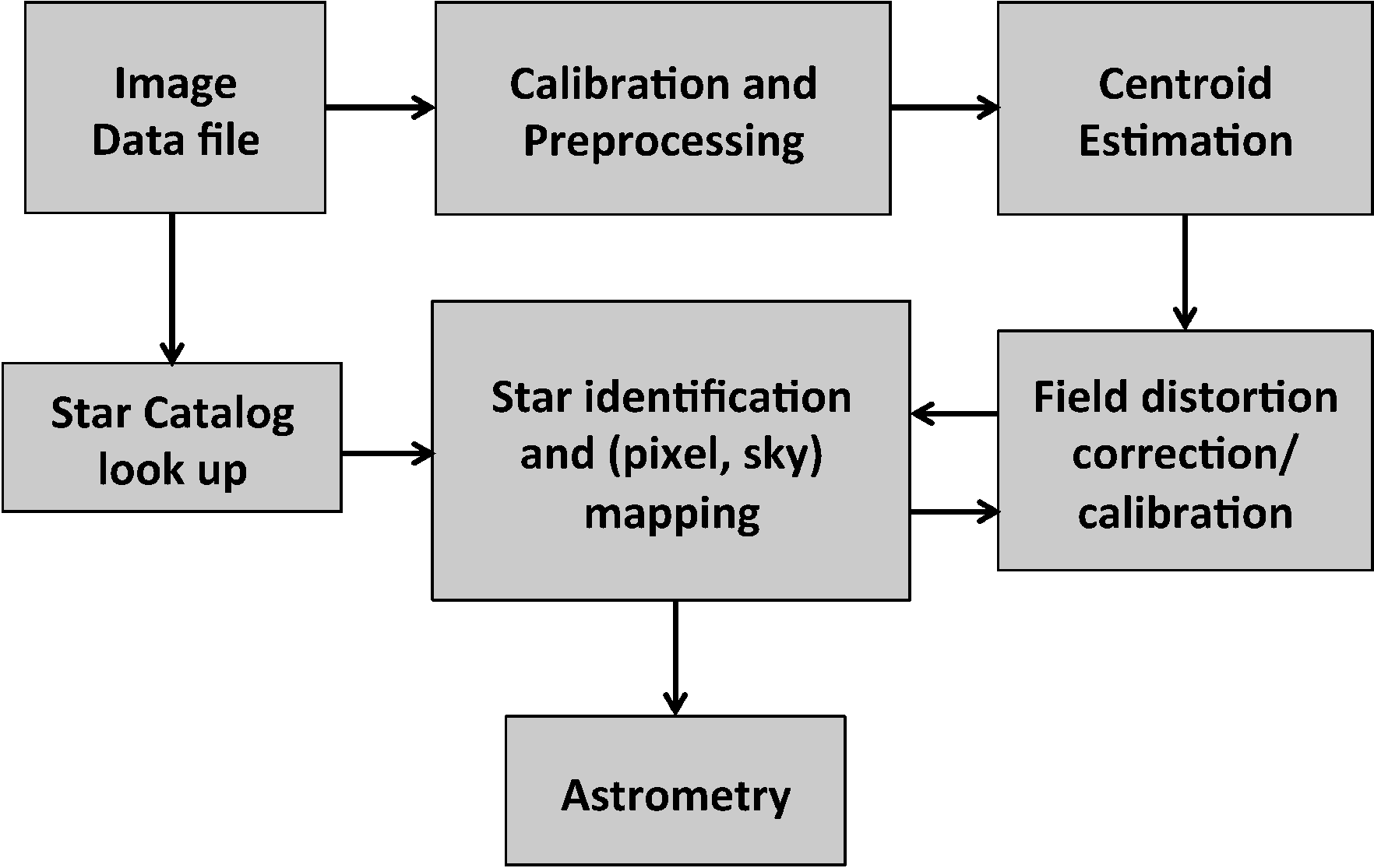}
\caption{A flow chart of the data processing for generating astrometry.\label{flowChart}}
\end{figure}

Fig.~\ref{flowChart} provides an overview of the data processing. 
We start with the raw images and apply calibration data to subtract the dark frame and factor out the flat field responses estimated using twilight images. 
We then perform preprocessing to remove the cosmic ray events and bad pixel signals.
Cosmic ray events are identified as signal spikes above random noise level localized in both temporal and spatial dimension.
The least-squares fitting is used to estimate centroid of reference stars.
A pre-estimated field distortion correction, modeled as low order polynomial functions, is applied to the reference stars in pixel coordinates.
The telescope pointing and the size of FOV are used to look up star catalog, the Gaia Data Release 1 (DR1) catalog.
 A planar triangle matching algorithm identifies stars in the field by matching congruent triangles from the field and the catalog with
 the shape determined by the relative distances between the stars.
 A set of identified stars enables us to solve for an affine mapping between the pixel coordinate and the position in the plane of sky, and thus the right ascension (RA) and declination (DEC).
 With this mapping, we can covert the pixel coordinate of the asteroid (after the field distortion correction) to RA and DEC in sky.

\subsection{Centroiding with synthetic tracking}
\label{sec:centroiding}
For bright objects, it is possible to estimate their centroid using single short-exposure images.
We thus can do NEA astrometry with these centroids from individual frames similar to stellar astrometry because
both moving target and the reference stars have compact PSFs in these short exposure frames.
However, in most of the cases, we would need to use the whole data cube to estimate the centroid
because when the signal in each short exposure frame is low, centroiding may not be reliable.
Instead, we do a least-squares fitting of ``moving point-spread-function'' to the whole data cube using
the following cost function
\beq
  C(v_x, v_y, x_c, y_c, \alpha, I_0 ) \equiv \sum_{x,y,t} \left | I (x, y, t) -\alpha P \left (x {-}X(t), y {-}Y(t) \right )  - I_0 \right |^2 w(x, y, t)
\eeq
where we parameterize a moving PSF, $P \left (x{-}X(t), y{-}Y(t) \right )$, with $P(x,y)$ being the PSF function
and $\left (X(t), Y(t) \right )$ represent the location of the object in frame $t$. We use a Moffat function \citep{Moffat1969} to model the PSF and determine the parameter
using a bright star in the field. The Moffat PSF function includes Gaussian PSF as its special case and fits better than a Gaussian to a seeing limited PSF; 
but the difference between using a Moffat and Gaussian PSF is much less than 10 mas.
$w(x,y,t)$ is a weighting function and $N$ is the total number of frames.
To minimize the variance of the estimation, the weight can be chosen to be the inverse of the variance of the measured $I(x,y,t)$, including photon shot noise and
sky background, dark current and read noise according to the Gauss-Markov theorem \citep{Luenberger1969}. We typically choose $w=1$ for simplicity because
the noise in $I(x,y,t)$ usually is not a limiting factor.
For most of the objects, the motion is linear and can be modeled as
\beq
   X(t) = x_c + v_x \left (t - (N{+}1)/2 \right ) + \epsilon_x (t) \,, Y(t) = y_c + v_y \left (t - (N{+}1)/2 \right ) + \epsilon_y (t)
\eeq
where $(x_c, y_c)$ is the location of the object at the center of the integration time interval and $(v_x, v_y)$ is the velocity of the linear motion.
$(\epsilon_x (t), \epsilon_y(t))$ is the tracking error with respect to sidereal, which can be estimated, for example, by an average of the centroids 
of a few bright reference stars in each frame. The location $(x_c, y_c)$ and velocity $(v_x, v_y)$ are solved simultaneously using a least-squares fitting
together with parameters $\alpha$ and $I_0$, which give photometry and background intensity.
We note that this approach to integrate a data cube avoids streaked images and keeps the jitter effect from atmosphere and telescope pointing common between
target and reference objects, thus achieves accuracy comparable with stellar astrometry, which we shall present in subsection~\ref{sec:centAccuracy}.

\subsection{Astrometric solution}
After the star identification using a triangular matching scheme, we have a mapping between pixel coordinates of the $N_s$ stars
$(X, Y)_i, i = 1, 2, \cdots, N_s$ and the catalog positions, the RAs and DECs, $(RA, DEC)_i,  i = 1, 2, \cdots, N_s$. 
We first convert the sky positions $(RA, DEC)_i$ into positions $(\xi, \eta)_i$ in the tangent plane and then solve for the following fitting
\beqa
   a \xi_i + b \eta_i + X_0 & \!\!\!=&\!\!\! X_i - \sum_{n,m, 2 \le n{+}m \le N_d} C^X_{n,m} X_i^n Y_i^m
\\
   c \xi_i + d \eta_i + Y_0 &\!\!\! =& Y_i - \sum_{n,m, 2 \le n{+}m \le N_d} C^Y_{n,m} X_i^n Y_i^m
\eeqa
where $N_d$ is the order of polynomial, which we found it is sufficient to have $N_d = 5$ for mas level calibration
and $a,b,c,d$ defines the linear transform between the sky plane position and measurements in pixel coordinates and $C^X_{n,m}, C^Y_{n,m}$ are distortion model coefficients.
$(X_0, Y_0)$ is an offset between the origins of tangent plane and the origin of pixel coordinate, which would be zero if the telescope had zero pointing error.
To improve the fitting accuracy, we can weight the fitting according to the inverse of the variances of uncertainties of $X_i$ and $Y_i$ from the centroiding fitting,
which is effective when faint reference stars are included.

\section{Results}
In this section, we present results from our instrument on the Pomona 40 inch telescope at the TMF. We first show the
centroiding results to have expected accuracy and then the overall performance of the astrometry with respect to the ephemeris from the JPL Horizon System.
\subsection{Astrometric Precision using Synthetic Tracking}
\label{sec:centAccuracy}
Synthetic tracking avoids streaked images by having exposures short enough so that the moving object does not streak in individual images,%
\footnote{More precisely, the length of streak is much smaller than the size of the PSF, it can be shown that the first order effect is proportional to the square of the ratio of streak length over PSF size.}
allowing us to achieve astrometry similar to stellar astrometry for NEAs.

\begin{figure}[ht]
\epsscale{0.8}
\plotone{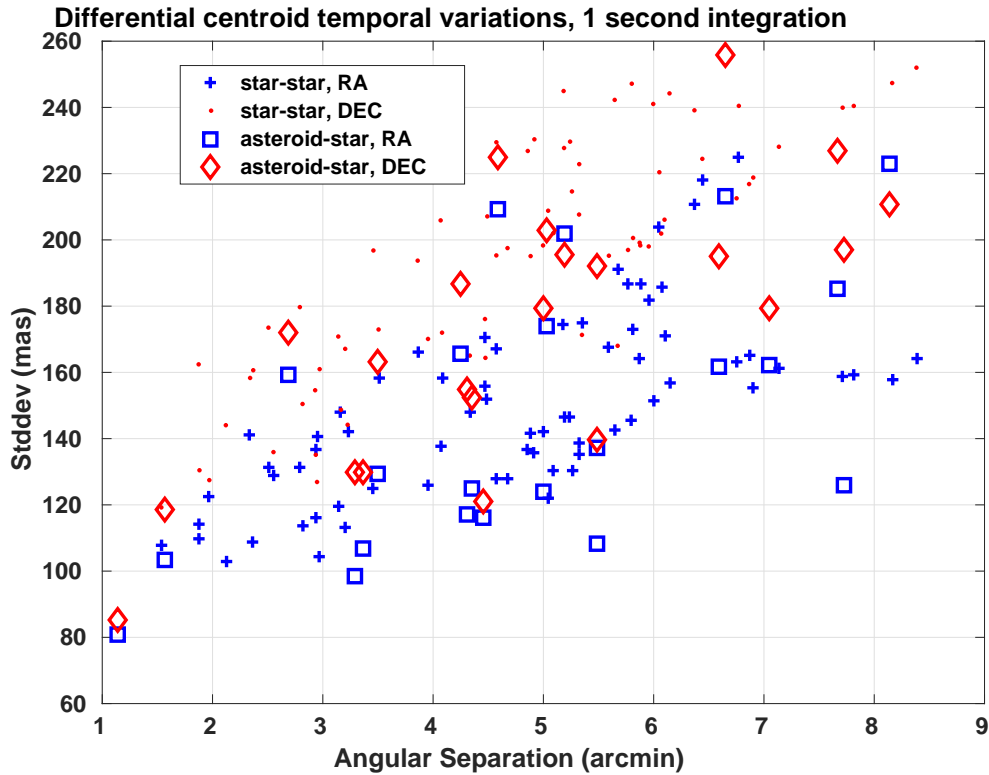}
\caption{Frame-to-frame standard deviations of differential centroid between asteroid and reference stars as function of the angular separation.\label{deltaCentroid}}
\end{figure}

To illustrate this, we estimate the centroid positions of a bright (apparent magnitude of $\sim$13.2) asteroid 1983 TB, observed on Dec 20, 2017,
 at distance of $\sim$0.09 AU from the Earth,  with respect to reference stars in each of the 1Hz frames.
 Fig.~\ref{deltaCentroid} shows the frame-to-frame standard deviations of the differential centroids between stars (blue plus sign for RA and red dot for DEC respectively) as function of the angular distances.
The frame-to-frame standard deviations serve as a measure of random errors for integration of 1 second, the exposure time.
Similarly, we plot also the frame-to-frame standard deviations of the differential centroid of asteroid 1983 TB with respect to reference stars in the same field 
after removing a linear motion of (0.596,-0.599) pixel/second consistent with JPL Horizon ephemeris for comparison,  marked with the blue squares and red diamonds representing RA and DEC respectively.
We can see that the random errors are comparable for the differential centroids between asteroid and stars and stellar differential centroids.
The random errors are mainly due to atmospheric turbulence because both the asteroid and reference stars are brighter than 16th magnitude, thus increasing
with the angular separation.

\begin{figure}[ht]
\epsscale{0.8}
\plotone{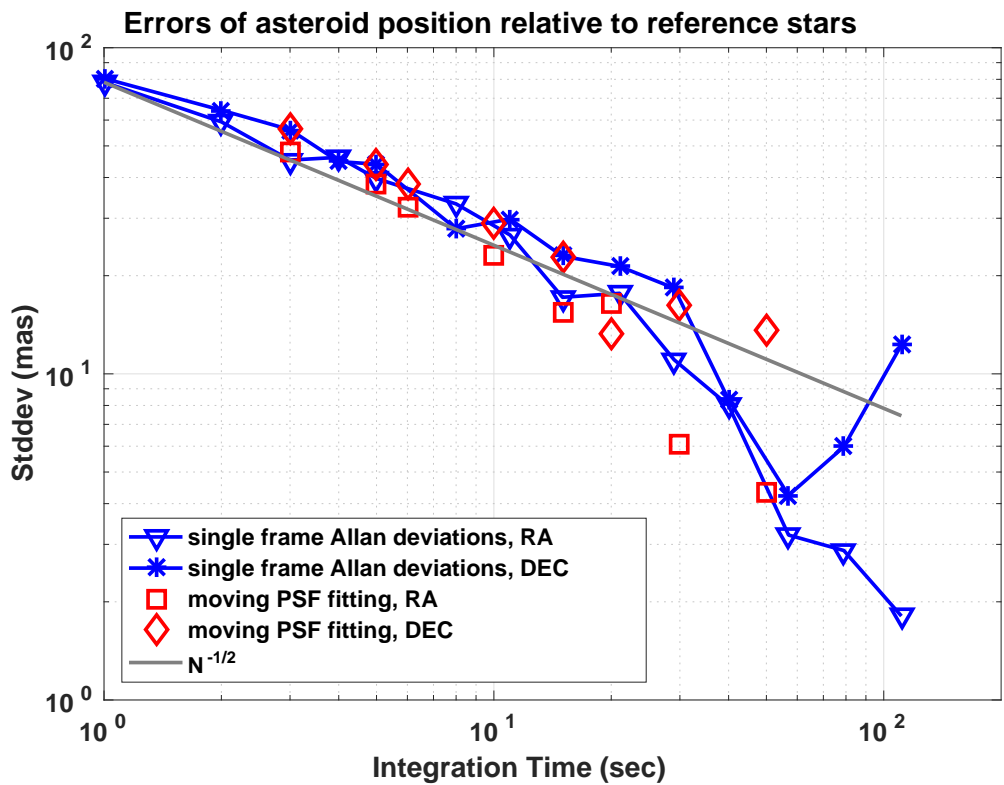}
\caption{Random error standard deviations as function of integration time. Blue lines with asterisk and triangle markers
represent the Allan deviations from integrating estimated differential centroid of asteroid 1983 TB with respect to a reference stars about 1 arc minute away.
The red squares and diamonds represent the estimated centroid using ``moving PSF fitting" to data cubes of the corresponding integration time.\label{movingPSFInt}}
\end{figure}

We in general estimate the centroid using the ``moving PSF fitting" to the data cube as described in the subsection~\ref{sec:centroiding}.
The ``moving PSF fitting'' to the data cube achieves essentially the same accuracy as averaging the estimated centroids from each short exposure frames.
Fig.~\ref{movingPSFInt} displays the Allan deviations for averaging centroid estimated from
each individual frames based on totally 300 frames after removing the linear motion of the asteroid
(blue lines with asterisk and triangle markers for RA, DEC respectively) as function of integration time.
Similarly, we can divide the 300 frames into a bunch of ``sub data cubes" corresponding to a specific integration time. For example, we have 30 ``sub data cubes" for 10 second
integration with each ``sub data cube" containing 10 frames. We apply ``moving PSF fitting" to each of the 30 ``sub data cube" to obtain 30 centroid estimates.
We can then estimate the centroiding error as the standard deviation of the 30 centroid estimates after removing the linear motion of the asteroid. 
The red squares and diamonds in Fig.~\ref{movingPSFInt} represent RA and DEC centroiding errors using ``moving PSF fitting" for the corresponding integration time.
The integration shows the expected the inverse of the square root of integration time behavior (gray line) because the errors are uncorrelated noises.
The ``moving PSF fitting'' gives a similar performance to estimates from averaging centroids estimated using individual frames.
Because using individual short exposure frames, the centroiding of asteroids and stars has similar performance as shown in Fig.~\ref{deltaCentroid},
and the integration down follows the inverse of square root of integration time behavior, synthetic tracking yields NEA astrometry with accuracy similar to that of stellar astrometry.
The ``moving PSF fitting" to the whole data cube is useful when the asteroid is dim
because centroiding individual short exposure frames may become too noisy to converge reliably.

The situation is, however, very different in the case when we estimate centroid using streaked images.
\begin{figure}[ht]
\epsscale{0.95}
\plotone{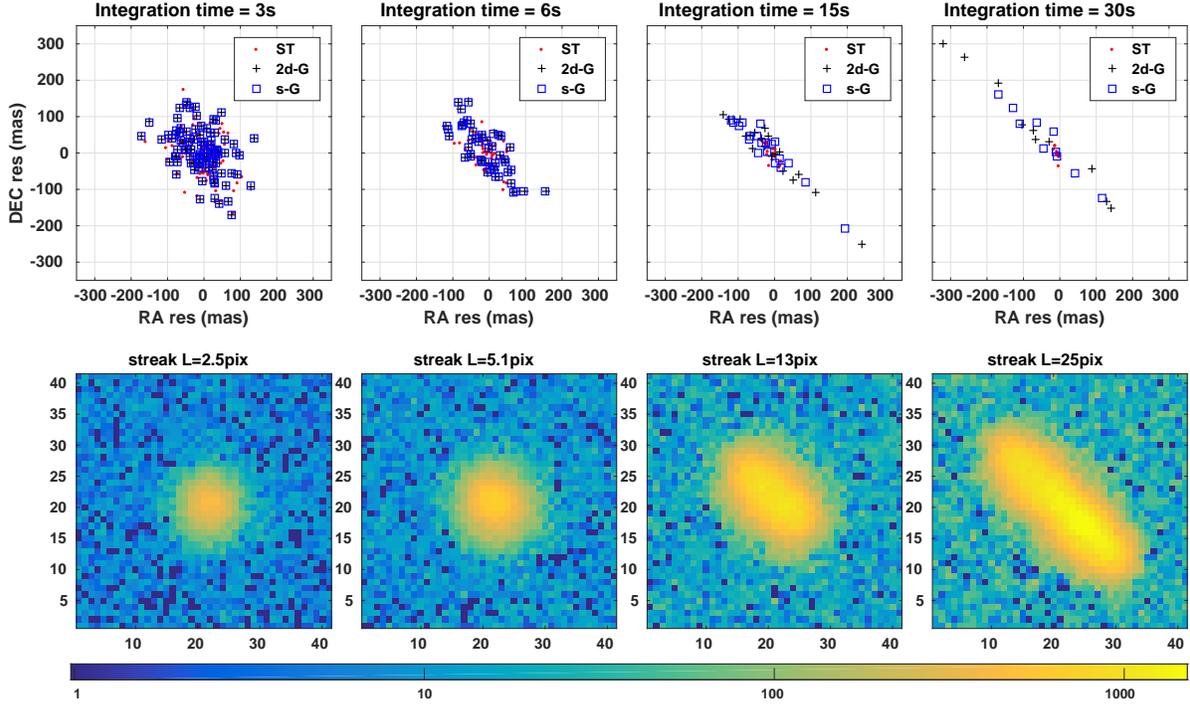}
\caption{Residuals (top) of differential centroid after removing linear motion
using three different centroiding methods (ST = synthetic tracking, 2d-G = two-dimensional elliptical Gaussian PSF fitting, s-G = streaked-Gaussian PSF fitting) using streaked images (bottom).  We can clearly see the degraded performance with the increase of integration time or the streak length. Note that the images are displayed in log scale, where the negative values are displayed according to the their absolute values to show the noise level.\label{resCmp}}
\end{figure}
For comparison, we combine images to simulate longer exposure images, and display the differential centroid residuals of asteroid 1983 TB with respect to
a nearby reference star after removing the same linear motion.
The performance of astrometry from centroiding streaked images clearly shows degradation with respect to synthetic tracking, especially along the streak as shown
in Fig.~\ref{resCmp}, where we display residuals from using synthetic tracking (ST, red dots), fitting with a two-dimensional elliptical Gaussian PSF (2d-G, black cross), and
fitting with a streaked-Gaussian PSF (s-G, blue squares) together with the corresponding 1983 TB images of different streak lengths.
As we increase integrate time from 3 s to 6 s, the spread of all the residuals shrinks. Further increasing integration time, the performance along streak
degrades due to the fact that the jitter effects from atmosphere and telescope pointing are no longer common between 
the asteroid and the nearby reference star. This may be understood as the centroid position of a streak along the streak is mainly determined
by the signals at both ends and is insensitive to the central portion of the image. Therefore, only the jitter effects at the beginning and end of
the integration affect the centroiding along the streak while for a well tracked reference star, the jitter effects during the whole integration affect the centroiding
of the reference star.

\begin{figure}[ht]
\epsscale{1}
\plottwo{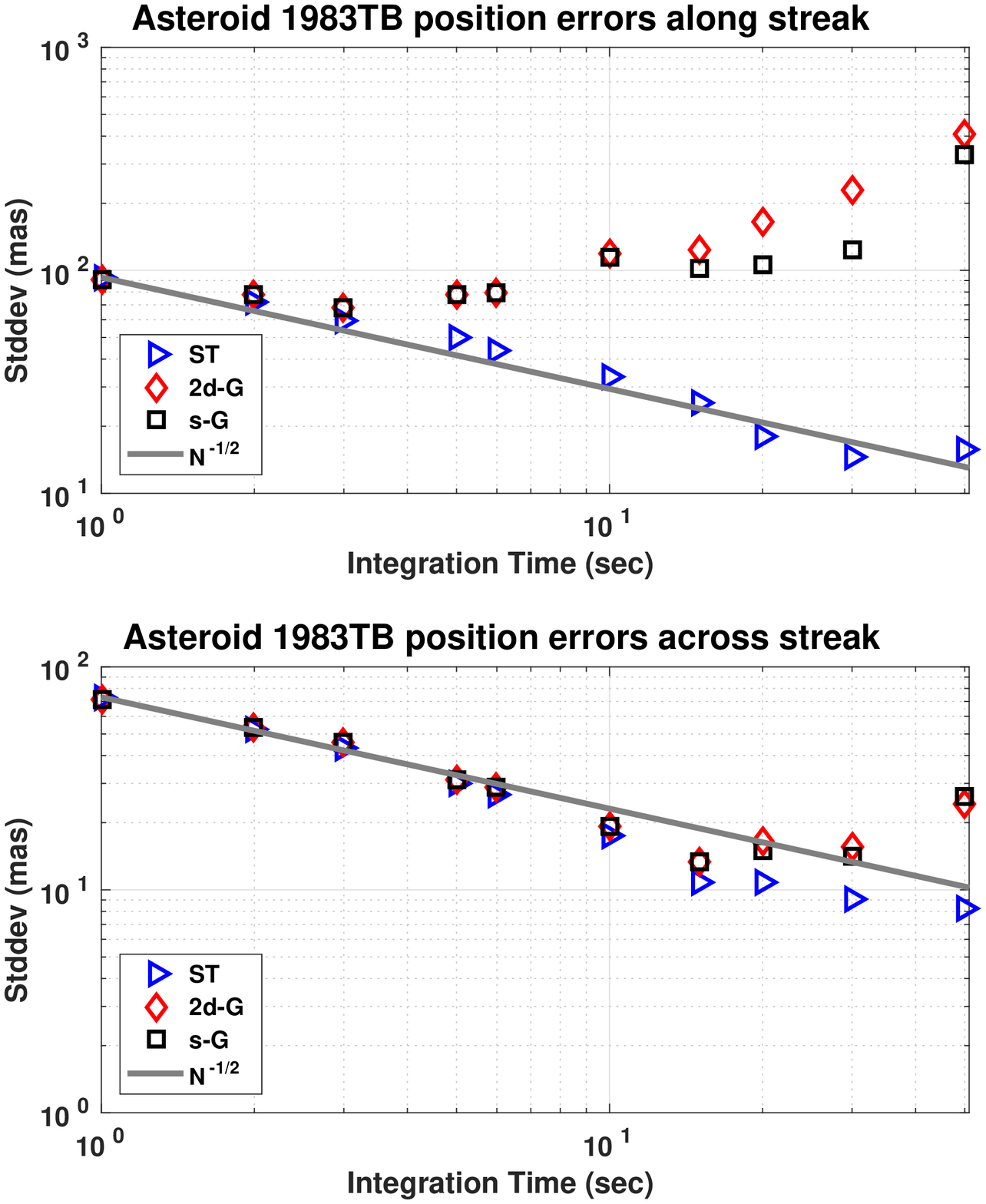}{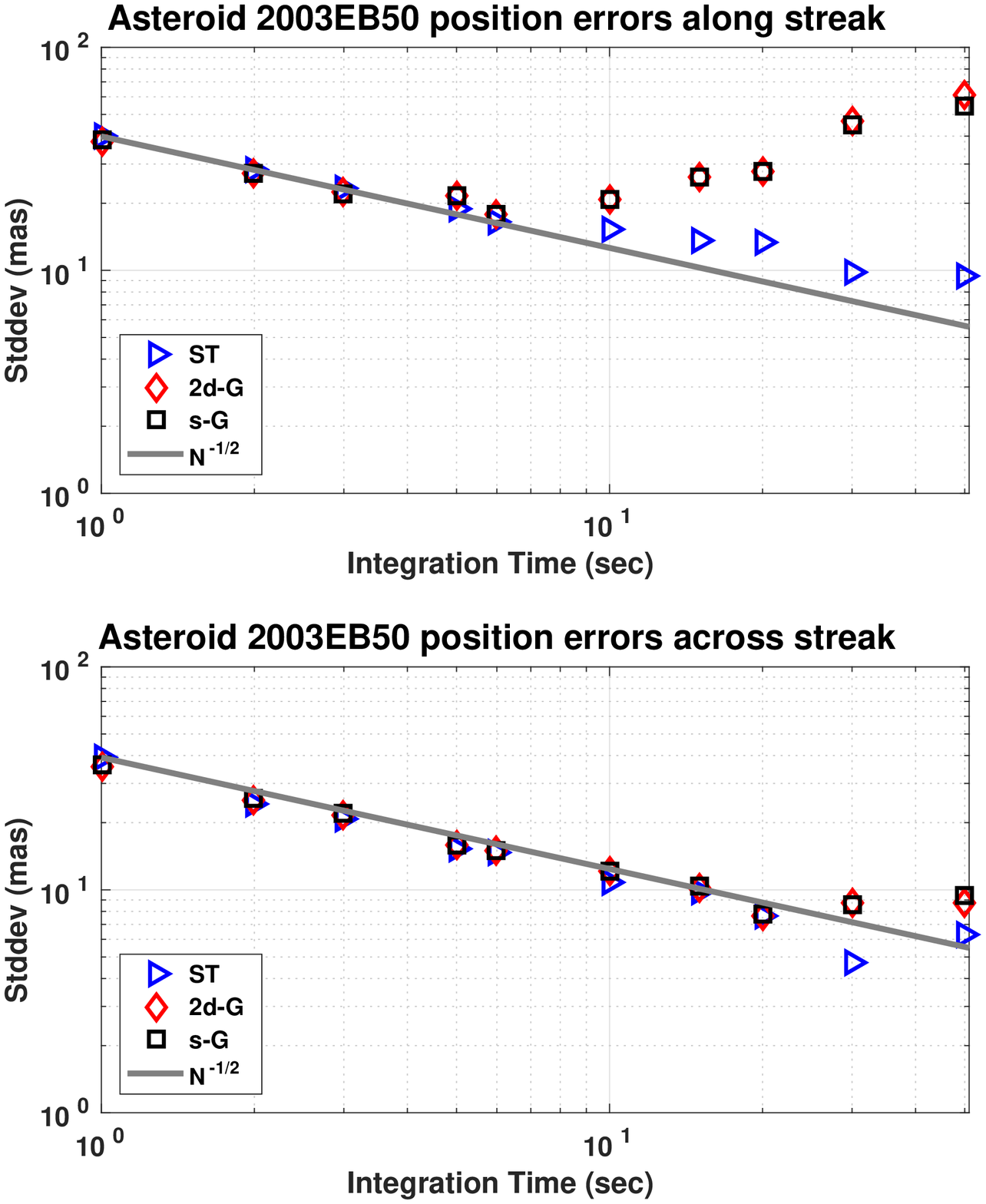}
\caption{Standard deviations of differential centroid between asteroid and a reference star using synthetic tracking and
2-d elliptical Gaussian PSF fitting, and streaked-Gaussian PSF fitting for asteroids 1983 TB (left) and 2003 EB50 (right).\label{centroidStdCmp}}
\end{figure}

Fig.~\ref{centroidStdCmp} shows centroiding performance along and across the streak
using a 2-d elliptical Gaussian PSF, streaked Gaussian PSF \cite{Veres2012}, compared with synthetic tracking method for asteroids 1983 TB (left) and 2003 EB50 (right)
respectively.  Asteroid 1983 TB moves at $\sim 0.38 \as$ per second, so it takes about 5 seconds for the streak length to be comparable with size of the PSF, 
at which the precision of centroiding using streaked images starts to degrade significantly. Asteroid 2003 EB50 moves at $\sim 0.1 \as$ per second, 
so it takes more than 10 seconds to see the degradation of accuracy.
At 1 Hz, errors are dominated by the air turbulences because the asteroids and reference stars are all bright.
The reference star for 2003 EB50 is $\sim 25\as$ away and the reference star for 1983 TB is $\sim 68 \as$ from the asteroid.
A closer reference star is the reason why the error at integration time of 1 second is smaller for 2003 EB50.
As expected, for long steaks, streaked-Gaussian PSF fitting gives the better performance than a 2-d elliptical Gaussian PSF fitting\citep{Veres2012}
for centroiding along the streak because the streaked-Gaussian PSF is a higher fidelity model of the intensity distribution of a streak.
Across the streak, the atmospheric effect is common between the asteroid and reference star, 
therefore, the performances are comparable until the degraded SNR from the increased sky background starts to hurt the performance.
For asteroid 1983 TB, the sky background is $\sim$5 photon per second per pixel.
As we integrated the air turbulence down to less than 15 mas at about 30 seconds, the sky background is 150 photon, becoming
comparable with the average pixel intensity of the asteroid signal, which is a few hundreds of photons.
Therefore, sky background noise starts to degrade the SNR, thus increase the the centroiding error. 
Similarly, this is also true for asteroid 2003 EB50. The degradation of SNR due to extra sky background noise happens at about 50 second integration
because it moves slower.

\subsection{Asteroid Astrometry}
We now present asteroid astrometry from observations taken since June 2017. 
Gaia DR1  \citep{Gaia2016} is used for our data reduction, but it does not have proper motions except for the Tycho stars.
We estimate the proper motion for the stars that are in both Gaia DR 1 and UCAC4 (\url{http://ad.usno.navy.mil/ucac/readme\_u4v5}) catalogs,
assuming a linear motion between epoch J2015 (Gaia ) 
and J2000 (UCAC4) . Because the accuracy of UCAC4 is about $\sim$50 mas, so our proper motion is
only accurate to a few mas per year. Propagating from Gaia's 2015 astrometry to about 2.5 years into 2017, we have a propagation
error generally less than 10 mas.

\begin{figure}[ht]
\epsscale{0.35}
\plotone{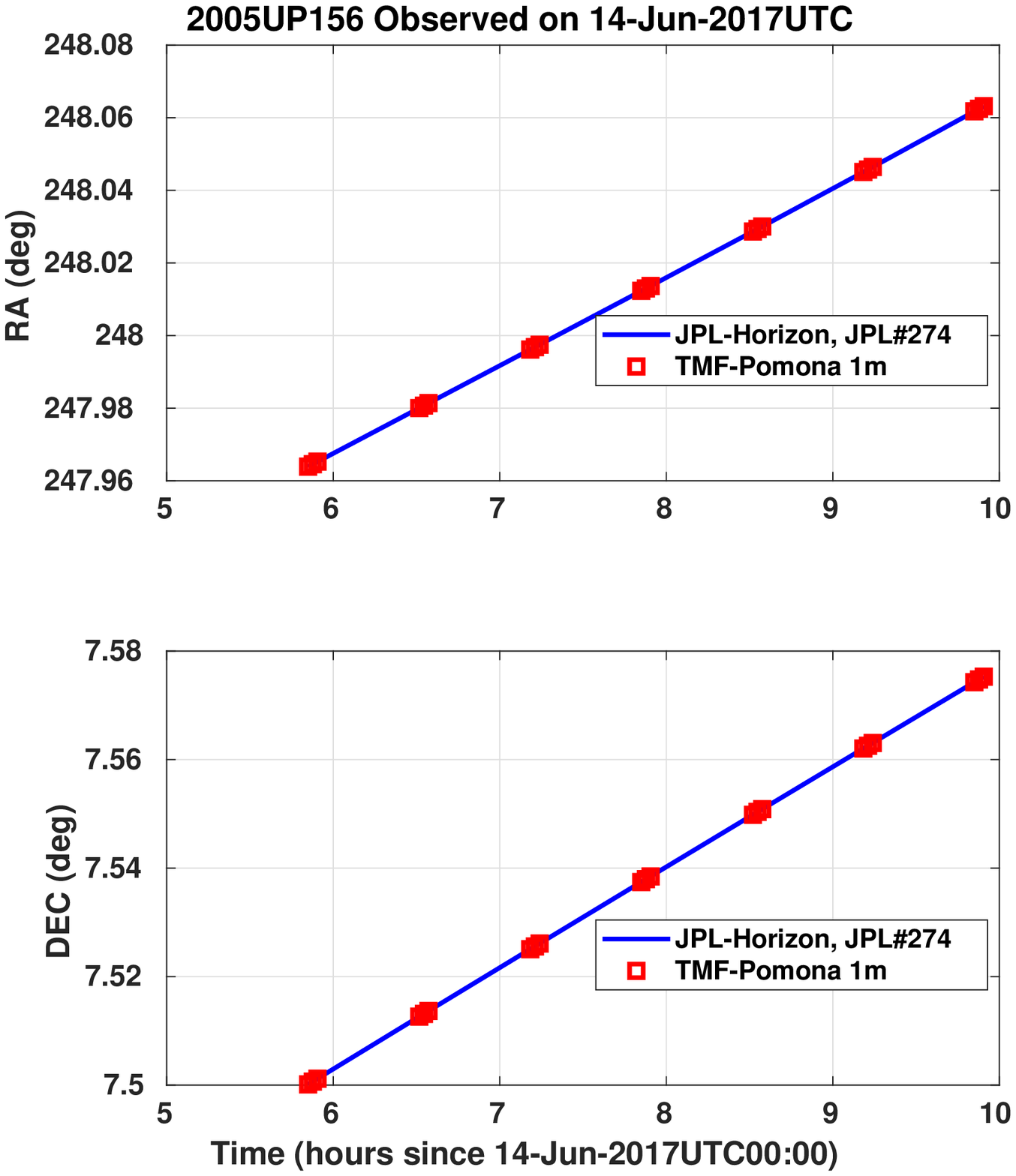}
\epsscale{0.57}
\plotone{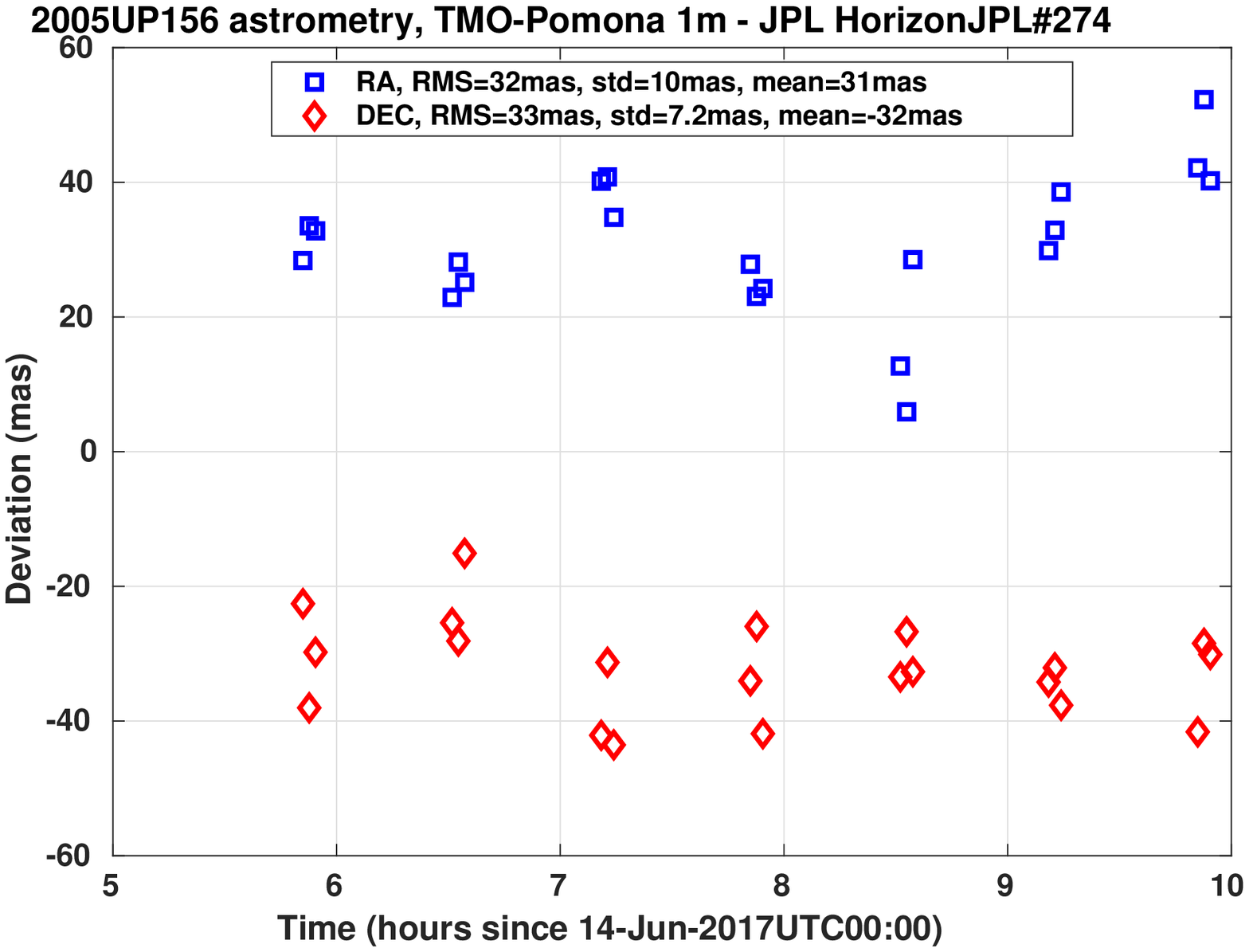}
\caption{Astrometry residuals of 2005 UP156 observed at TMO after subtracting the JPL Horizon System solution \#274.\label{AstrometryHoursSlow}}
\end{figure}
Fig.~\ref{AstrometryHoursSlow} shows astrometry of asteroid 2005 UP156 observed on June 14, 2017 with the left plot displays respectively the RA and DEC 
from our instrument  (red squares) on top of the ephemeris from the JPL Horizon System (\url{https://ssd.jpl.nasa.gov/?horizons}) (blue curve). The right plot displays
the residuals, the difference between our astrometry and JPL Horizon system solution \#274. Each of the data point is derived from integrating
100$\times$1Hz frames (100 seconds). The spread of our measurements have standard deviations of $\sim$10 mas.
There is clear bias around 30 mas, which we believe is due to the uncertainty of the JPL Horizon System solution $\sim$100 mas at 3-$\sigma$ level.
2005 UP156 is a kilometer size asteroid and was at $\sim$0.16 AU from earth during the observation with an apparent magnitude of $\sim$14.7, moving at a slow speed of
$\sim$2$\as$ per minute.

\begin{figure}[ht]
\epsscale{0.35}
\plotone{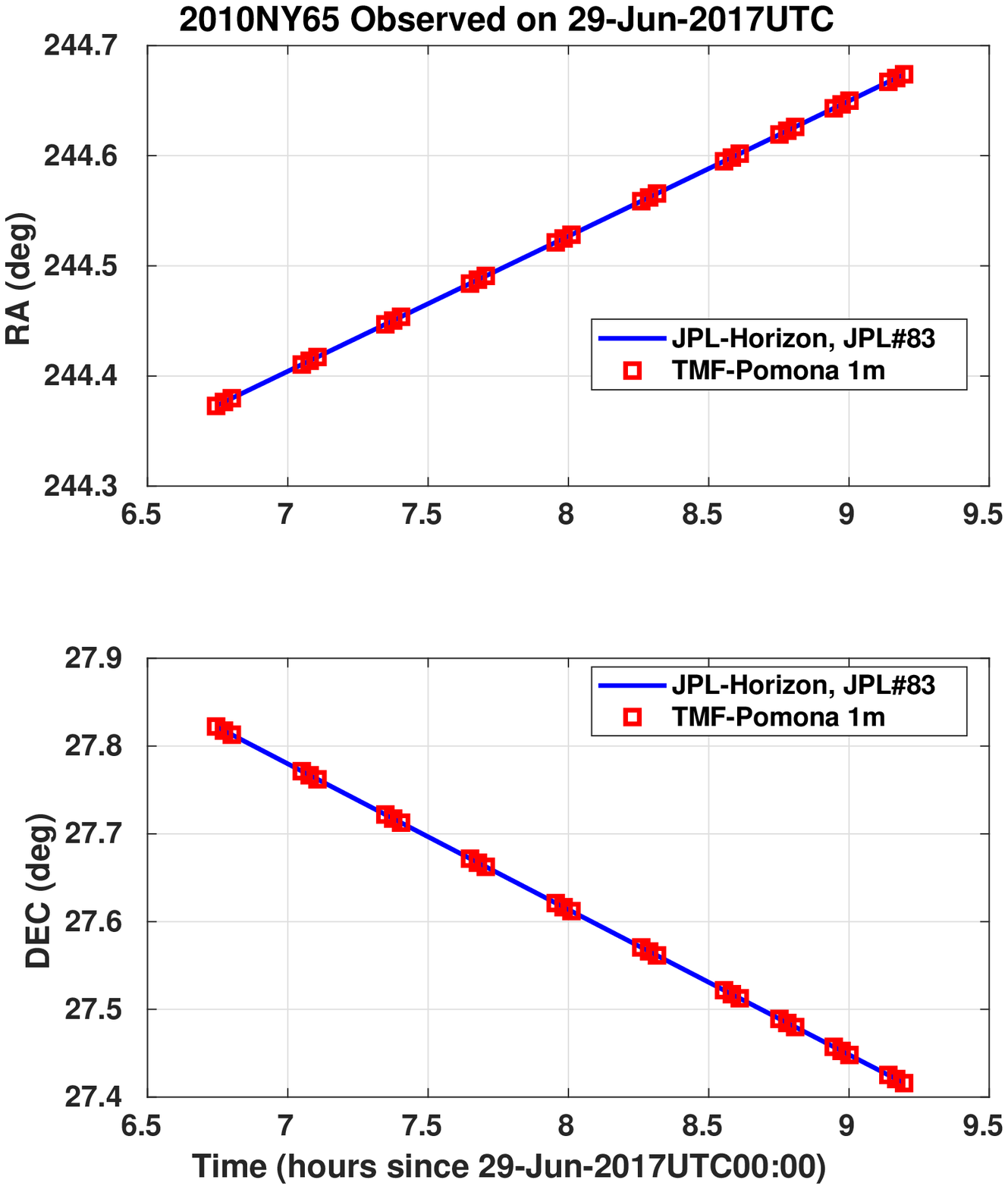}
\epsscale{0.57}
\plotone{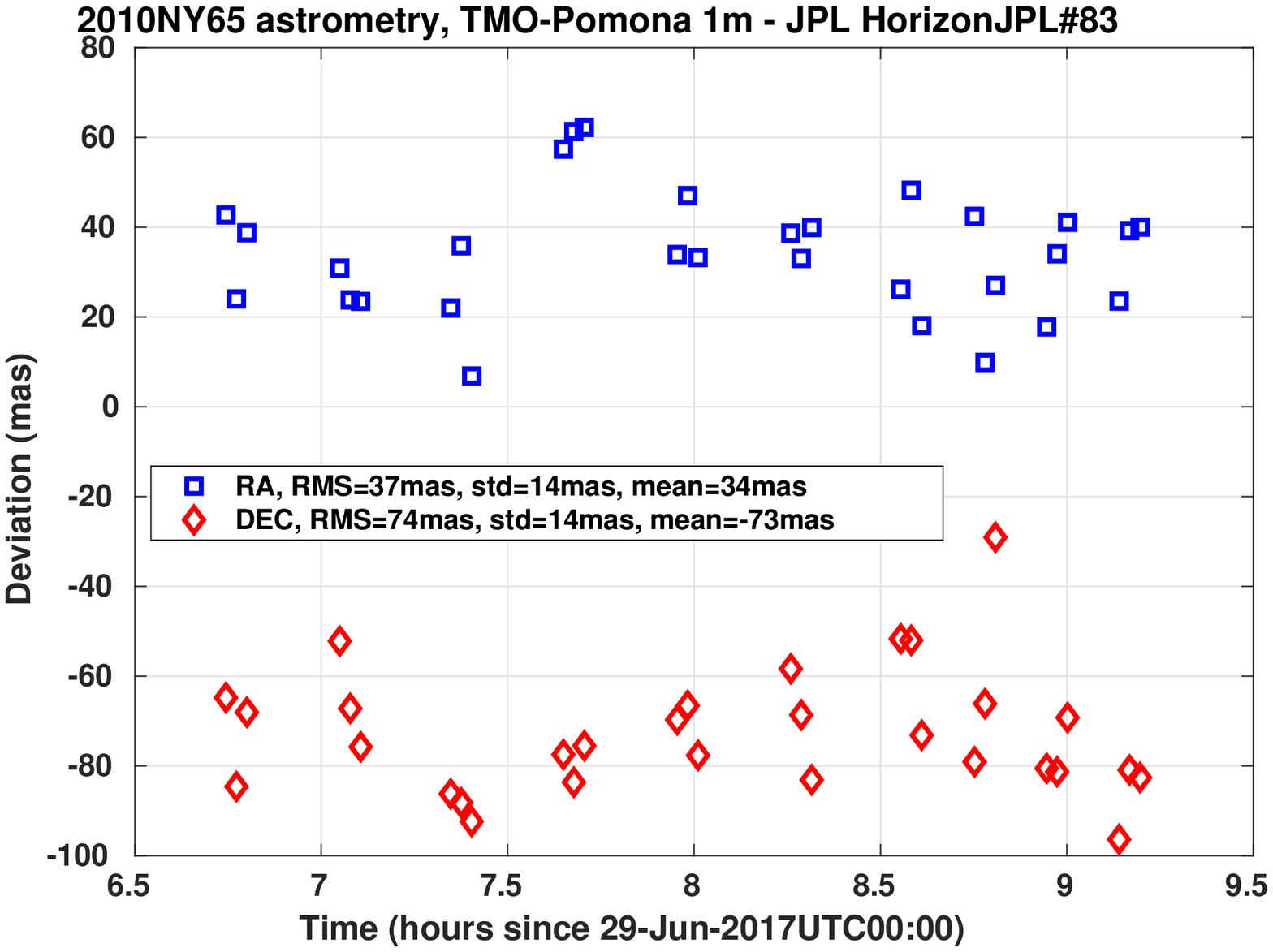}
\caption{Astrometry residuals of 2010 NY65 observed at TMO after subtracting the JPL Horizon System solution \#83.\label{AstrometryHoursFast}}
\end{figure}

\begin{figure}[h]
\epsscale{0.8}
\plotone{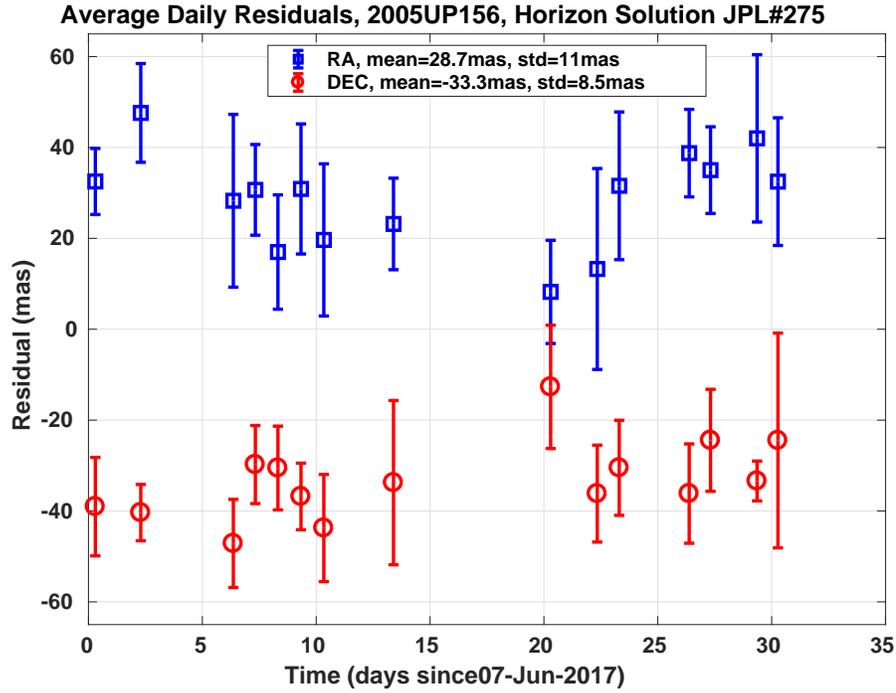}
\caption{Astrometry residuals of 2005 UP156 observed at TMO after subtracting the JPL Horizon System solution \#274.\label{AstrometryDays}}
\end{figure}

Fig.~\ref{AstrometryHoursFast} shows the similar plots for a faster moving asteroid, 2010 NY65 observed on June 29, 2017. This asteroid has a size $\sim$200 m
and was at $\sim$0.04AU away from the Earth during the observation with apparent magnitude $\sim$16.8. It moves at speed of 0.2$\as$ per second.
Again, the standard deviations (spread of our measurements) are $\sim$14 mas and there are biases consistent with the JPL Horizon System solution uncertainty
of solution \#83. Each of the measurements was obtained from integrating 100$\times$1Hz frames.
It is possible to further average down random errors to have better precision, however, we are limited by systematic errors. 
Fig.~\ref{AstrometryDays} displays astrometry residuals of 2005 UP156 from observations over five weeks, where  each data point represents the
average value of all the observations within one night with the standard deviations of data within a night shown as the error bar. 
The spread of daily mean astrometry however, is still $\sim$10 mas, not smaller from averaging data over one night; this suggests systematic errors.
The consistency of the overall constant bias over five weeks is unlikely due to our systematic errors; it is likely to be the prediction uncertainty of the
JPL Horizon system, whose 3-$\sigma$ is $\sim$ 100 mas.

It is usually sufficient to use only 3rd order polynomials to correct the field distortion to 10 mas over the 19$^\prime{\times}16^\prime$ field.
Our major error comes from chromatic distortion effect.
A typical field distortion calibration has 40-50 mas RMS of errors with 50 -200 reference stars in the field. 
Near the center of the field over a region about $6^\prime{\times}6^\prime$, we can correct to 10-20 mas.
In case of a wide range of stellar spectra, our field distortion over the whole field of view can be as large as $\sim$100 mas.
Currently we mitigate this effect by restricting reference stars according to colors, {\it e.g.}~using
the difference of magnitudes in B and V bands. 
Applying an R or V bandpass filter ($\sim$ 120nm)  helps us reduce field distortion calibration error to less than 20 mas
over the whole 19$^\prime{\times}16^\prime$ field, and better than 10 mas over the center $6^\prime{\times}6^\prime$ region.

\begin{table}[H]
\centering
\caption{NEA Observed\label{Table1}}
\begin{center}
\begin{tabular}{|c|c|c|c|}
\hline
Asteroid  & Date Range of Observation & Number of Observations  \\ \hline \hline
1984 KB & 20170607-20170707& 87 \\ \hline
1999 KW4 & 20170607-20170627& 47 \\ \hline
2000 PD3 & 20170731-20170814& 14 \\ \hline
2002 VU94 & 20170807-20170814 & 10 \\ \hline
2004 BG121& 20170706-20170617 & 8 \\ \hline
2004JB12 & 20170615-20170620& 10 \\ \hline
2005 UP156 & 20170607-20170811 & 152  \\ \hline
2007 WV4 & 20170607-20170614 & 27 \\ \hline
2010 NY65& 20170627-20170706 & 47  \\ \hline
2010 VB1 & 20170620-20170620 & 2 \\ \hline
2014 YC15 & 20170731-20170811& 13 \\ \hline
2017 BM31 & 20170706-20170731& 24 \\ \hline
\end{tabular}
\end{center}
\end{table}

Since June 2017, we have observed a dozen asteroids listed in Table~\ref{Table1} and reported more than 400 data points to the Minor Planet Center (\url{https://minorplanetcenter.net/}). Each of
our data points represent an integration over 300 seconds of data. The root-mean-square (RMS) of our astrometry residuals is about 50 mas as shown
in Tabel~\ref{Table2}, where we have obtained the RMS of NEA astrometry residuals for Pan-STARRS, which has the smallest RMS among the major NEA surveying facilities,
from the reference \cite{Veres2017}.
Because we believe a major contribution to the RMS is the uncertainty of the predicted asteroid ephemeris from the JPL Horizon System
just like in case of Fig.~\ref{AstrometryDays}, we also compute
the standard deviations of our data points over each night just like what we did for the spread in Fig.~\ref{AstrometryHoursFast}., which we believe
is closer to our accuracy for most of the data points.
\begin{table}[H]
\centering
\caption{NEA Astrometry Residual Comparison\label{Table2}}
\begin{center}
\begin{tabular}{|c|c|c|c|}
\hline
  & Pan-STARRS1 RMS & Pomona 40 inch RMS & Pomona 40 inch Std. Dev. \\ \hline \hline
 RA (mas) & 120 & 51 & 29 \\ \hline
 DEC (mas) & 120  & 53 & 24 \\ \hline
\end{tabular}
\end{center}
\end{table}

\begin{figure}[ht]
\epsscale{0.4}
\plotone{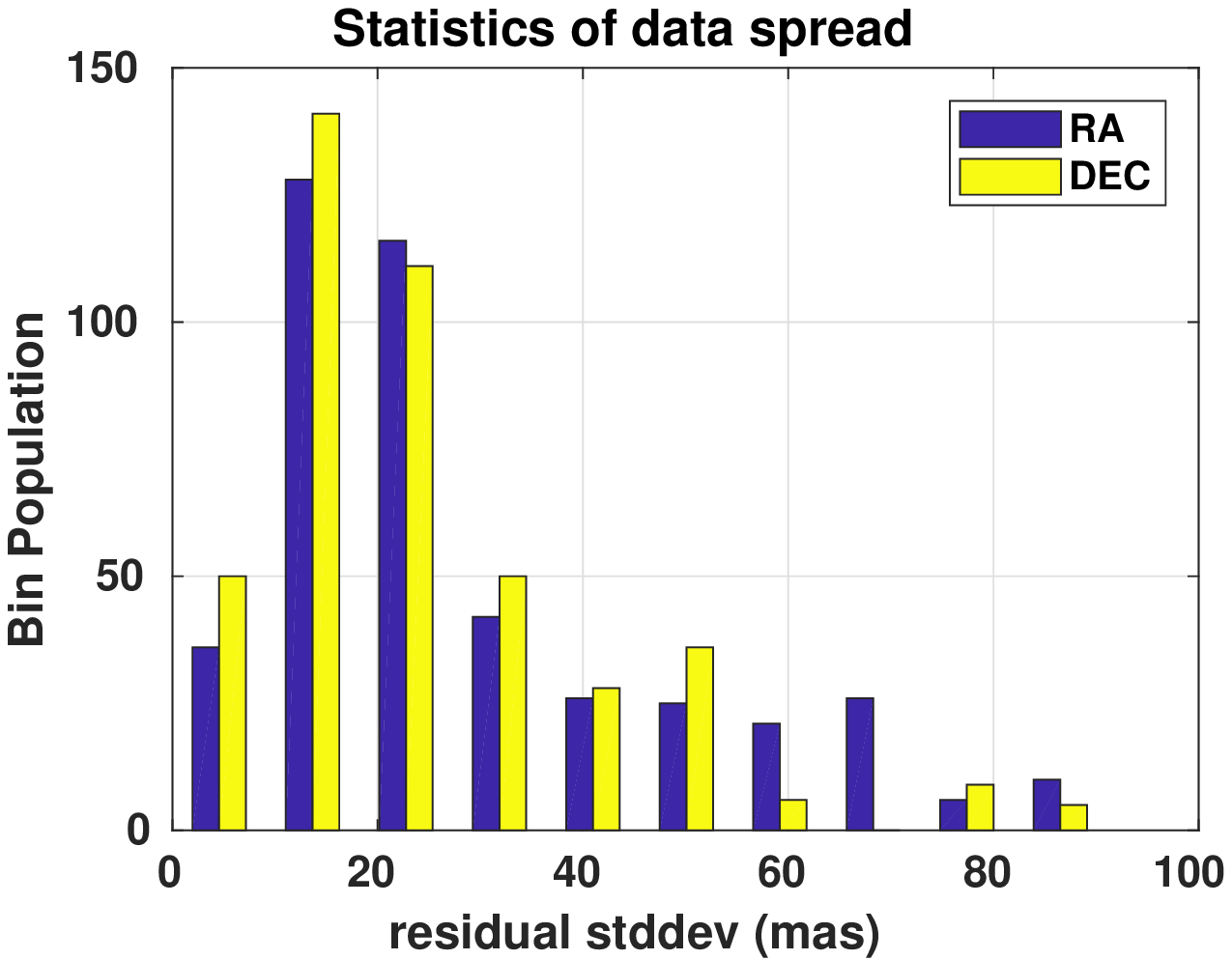}
\epsscale{0.47}
\plotone{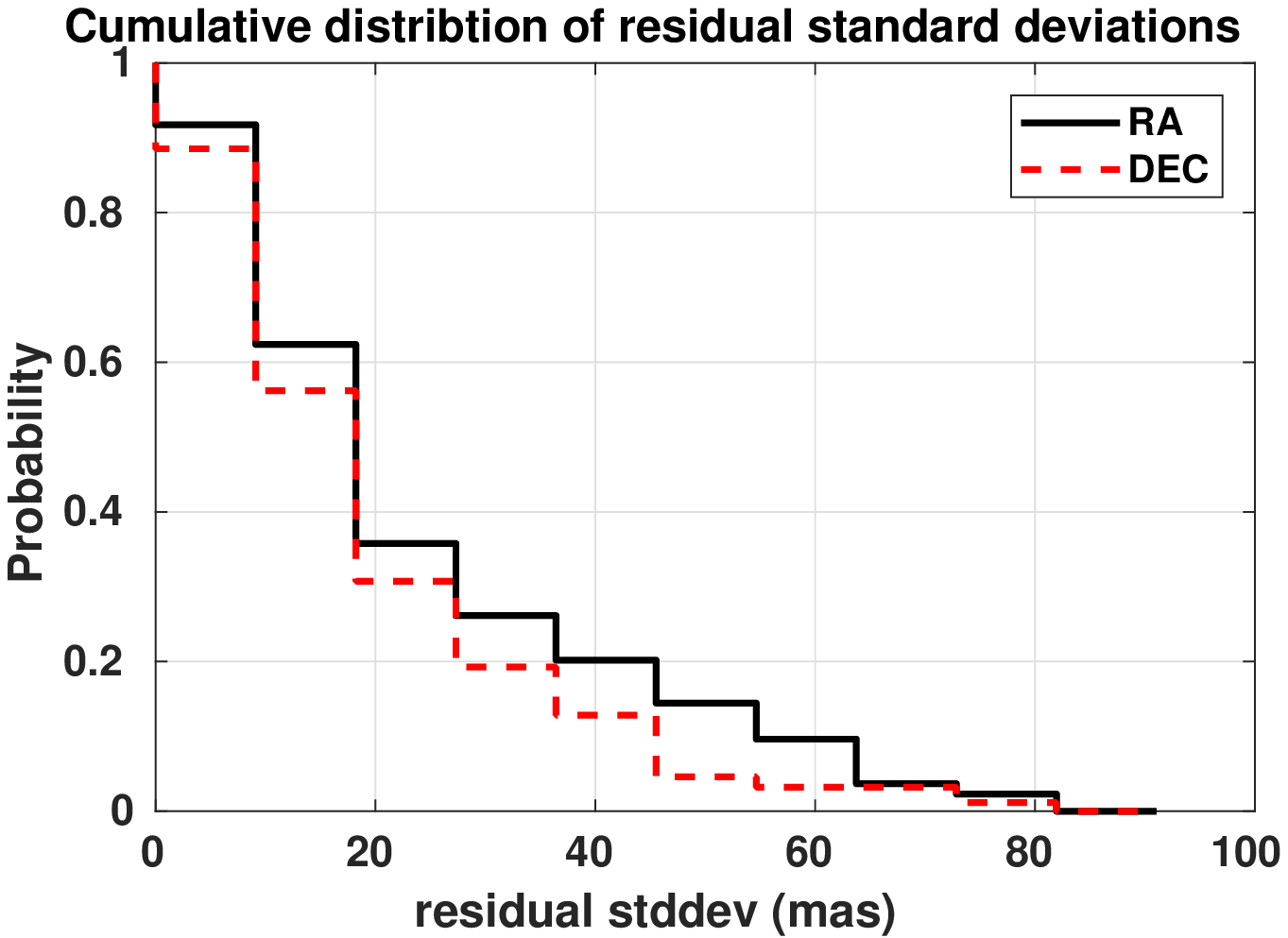}
\caption{Astrometry residuals statistics.\label{ResStat}}
\end{figure}

Left plot in Fig.~\ref{ResStat} displays the histogram of the standard deviations of our astrometry residuals of reported data within a night
and the right plot shows the corresponding cumulative distribution.
For some reason, the DEC has slightly better performance. More than 85\% of the data have accuracy better than 50 mas. 
The largest residuals can come from centroiding errors due to confusion field or random errors for faint targets.
For bright targets (mag $<$ 17), our accuracy are typically better than 20 mas with the best $\sim$10 mas.

Most of the Pan-STARRS astrometry statistics comes from slowly moving asteroids, probably main belt asteroids, with an integration time of 45 seconds.
According to Fig.~2 in reference \citep{Veres2017}, as the speed of asteroid motion increases, the degradation becomes clear.
In contrast, our accuracy is independent of how fast the target moves in sky with the synthetic tracking technique.

\section{Summary, Discussions, and Future Works}

In summary, using synthetic tracking technique to observe NEAs enables us to achieve astrometry with accuracy comparable with
stellar astrometry. We are able to achieve 50 mas level NEA astrometry with the data taken since June of 2017. Our best astrometry for bright targets
observed near the center of field can be at $\sim$10 mas with an integration time of 100 second. Our accuracy is insensitive to how fast the asteroid moves in the sky.

Our major source of systematic errors comes from the chromatic distortions due to refractive elements in the optics.
We are working along two threads to mitigate the effects: 1) applying narrow band filters; 2) putting a detector 
directly the at the Cassegrain focus to eliminate refractive optical elements in the system.
Because the chromatic effect is proportional to the square of the passband, with a 25nm passband, we are able to correction field distortion down to $\sim$8 mas.
Another source of systematic errors comes from the propagation of astrometry at Gaia's 2015 epoch to our observation epoch in 2017,
where we either assume the proper motion to be 0 or derive the proper motion by combining both UCAC4 and Gaia astrometry
for stars in both catalogs. With Gaia's Data Release 2 catalog, the proper motions and stellar colors will be available, so the
error due to unknown proper motion is expected to be less than 1 mas and the stellar color information can be used to correct
differential chromatic effect from the atmosphere.

The application of accurate ground-based astrometry to optical navigation for the future spacecraft that carry optical communication devices offers a convenient way
to operate. During the downlink, the ground terminal can take astrometric measurements simultaneously to determine the spacecraft position in the plane-of-sky.
Unlike current radio frequency (RF) approach of using the delta-Differential One-way Ranging  (delta-DOR) measurements based on deep space network (DSN) antennas,
whose measurement frequency is approximately once per day, we can take measurements more frequently. This could be also a solution for navigating future
deep space SmallSats, whose population is expected to grow fast, without overwhelming the existing DSN facilities.
The current state-of-the-art RF astrometry precision obtained from the delta-DOR measurements \citep{Border2004} has accuracy of 1-2 nrad (0.2-0.4 mas), which
is sufficient for the most stringent spacecraft navigation, such as orbit insertion.
Treating asteroids as proxy of these spacecraft, we are studying the feasibility of achieving the required $\sim$0.2 mas (1nrad)
accuracy for optical navigation using the Gaia's final catalog (DR 2 is not sufficient for 1nrad astrometry).

With more than 90\% of the asteroids larger than 1 km has been found, NASA NEO program is now searching for smaller asteroids down to 100 m or smaller.
The discovery rate of asteroids has been increasing reaching more than 2000 per year in 2017.
Looking into the future, we expect that high accuracy astrometry to play an important role to perform follow-up observations of these newly detected asteroids so as to 
provide measurements for initial orbit determinations.
Follow-up observations are crucial to keep the asteroids from being lost; efficiently doing so is extremely valuable to keep up with the increasing number of detections.
Theoretically, it is possible to derive an orbit from only three very accurate measurements, we thus believe only a few carefully scheduled follow-up observations
of high accuracy astrometry can be an efficient way of doing follow-up observations.
Future newly detected asteroids are likely to be smaller asteroids, therefore the detection will be at a closer distance to the earth, thus moving fast. 
Our approach of using synthetic tracking would be ideal to follow up these fast moving objects.

\acknowledgments
The authors would like to thank Heath Rhoades at the Table Mountain Facility of JPL for supporting the instrumentation
and Paul Chodas at JPL for technical advices on using the JPL Horizon System.
This work has made use of data from the European Space Agency (ESA)
mission {\it Gaia} (\url{https://www.cosmos.esa.int/gaia}), processed by
the {\it Gaia} Data Processing and Analysis Consortium (DPAC,
\url{https://www.cosmos.esa.int/web/gaia/dpac/consortium}). Funding
for the DPAC has been provided by national institutions, in particular
the institutions participating in the {\it Gaia} Multilateral Agreement.
The work described here was carried out at the Jet Propulsion Laboratory, California Institute of Technology,
under a contract with the National Aeronautics and Space Administration.
Copyright 2018. Government sponsorship acknowledged.

\clearpage

\end{document}